\documentclass[useAMS,usenatbib,aps,floatdix,nofootinbib]{mn2e}

\usepackage{color}

\usepackage{graphics}
\usepackage{amssymb,amsmath}
\usepackage{amsmath}
\usepackage{psfrag}
\usepackage{graphicx}

\newcommand{\f}{\frac}

\newcommand{\be}{\begin{equation}}      
\newcommand{\ee}{\end{equation}}      
      
\newcommand{\bef}{\begin{figure}}      
\newcommand{\eef}{\end{figure}}      
\newcommand{\bea}{\begin{eqnarray}}    
\newcommand{\eea}{\end{eqnarray}}      
  
\def\spose#1{\hbox to 0pt{#1\hss}}
\def\ltapprox{\mathrel{\spose{\lower 3pt\hbox{$\mathchar"218$}}
\raise 2.0pt\hbox{$\mathchar"13C$}}}
\def\gtapprox{\mathrel{\spose{\lower 3pt\hbox{$\mathchar"218$}}
\raise 2.0pt\hbox{$\mathchar"13E$}}}
\def\inapprox{\mathrel{\spose{\lower 3pt\hbox{$\mathchar"218$}}
\raise 2.0pt\hbox{$\mathchar"232$}}}

\def\bse{\begin{subequations}}
\def\ese{\end{subequations}}

\def\lsim{\raise 0.4ex\hbox{$<$}\kern -0.8em\lower 0.62ex\hbox{$\sim$}} 
\def\gsim{\raise 0.4ex\hbox{$>$}\kern -0.7em\lower 0.62ex\hbox{$\sim$}}

\def\f0N{f_0^{(N)}}
\def\bec{\begin{center}}
\def\eec{\end{center}}

\title[Universal properties of violently relaxed gravitational
  structures]{Universal properties of violently relaxed gravitational
  structures}

\author[F. Sylos
  Labini] {Francesco Sylos Labini${^{1,}}{^{2}}$\\ $^1$Centro Studi e
  Ricerche Enrico Fermi, Via Panisperna 00184 - Rome -
  Italy\\ $^2$Istituto dei sistemi complessi, Consiglio Nazionale
  delle Ricerche, Via dei Taurini 19, 00185 Rome, Italy\\ }
\begin{document}

\date{\today}

\maketitle

\begin{abstract} 
We study the collapse and virialization of an isolated spherical cloud
of self-gravitating particles initially at rest and characterised by a
power-law density profile, with exponent $0 \le \alpha < 3$, or by a
Plummer, an Hernquist, a NFW, a Gaussian profile.  We find that in all
cases the virialized structure formed after the collapse has a density
profile decaying, at large enough radii, as $\sim r^{-4}$, and a radial
velocity dispersion profile decaying as $\sim r^{-1}$.  We show that
these profiles originate from the physical mechanism responsible of
the ejection of a fraction of cloud's mass and energy during the
collapse and that this same mechanism washes out the dependence on the
initial conditions.  When a large enough initial velocity dispersion
is given to the cloud particles, ejection does not occur anymore and
consequently the virialized halo density and velocity profiles display
features which reflect the initial conditions.
\end{abstract}

\begin{keywords}
Galaxy: halo; Galaxy: formation; globular clusters: general;
(cosmology:) dark matter; (cosmology:) large-scale structure of
Universe; galaxies: formation
\end{keywords}


\section{Introduction}

It is well known that some of the properties of virialized structures
formed through the gravitational collapse of an isolated system depend
on the initial conditions in various ways; however, some other
statistical features of these structures seem to be generated by the
dynamical mechanisms acting during the relaxation process (see e.g.,
\cite{vanalbada_1982,aarseth_etal_1988,theis+spurzem_1999,boily+athanassoula_2006,paper2,paper3,visbal_2012}
and references therein).

For instance, when a spatially uniform and isolated cloud of density
$\rho_0$ and with a small enough initial velocity dispersion
collapses, it forms, in a time-scale $\tau_D=\sqrt{3\pi/(32G\rho_0})$,
a virialized structure characterised, for radii larger than the size
of a dense inner core $r_c$, by (i) a power-law density profile $n(r)
\sim r^{-4}$, (ii) a radial velocity dispersion which decays in a
Keplerian way $\sigma^2_r(r) \sim r^{-1}$ and (iii) a pseudo
phase-space density that decays as $n(r)/\sigma^3_r(r) \sim r^{-5/2}$
\citep{paper1}. 

These behaviors were shown to be associated with the ejection of a
fraction of the initial mass and energy from the system during the
collapse: in particular, particles with binding energies close to zero
follows almost radial orbits around the dense core, thus displaying a
Keplerian velocity dispersion. Considering that they conserve their
total energy it is possible to show that the density profile follows
the observed $n(r) \sim r^{-4}$ decay \citep{paper1}. Note that the
crucial role of particles ejection for the formation of non-trivial
density and velocity profiles has not been properly appreciated in the
literature (see discussion in \citet{paper2,paper1}).

In order to study whether the behaviors (i)-(iii) above are found in
the collapse of spherical clouds with different initial density
profiles than uniform and to determine whether in more general cases
there remains memory of the initial conditions after the collapse
phase, we consider (i) a set of clouds with an initial density profile
of the type $n(r) = {n_{\alpha} }/{r^\alpha} \;,$ where $n_\alpha$ is
a constant and $0\le \alpha < 3$ and (ii) density profiles without a
simple scale free behaviour, e.g., a Plummer sphere, an Hernquist
profile \citep{hernquist1990}, a Navarro Frenk and White (NFW) profile
\citep{navarro1996} and a Gaussian profile.

Note that while the formation and evolution of self-gravitating
dissipation-less systems is a theoretical interesting and complex
problem, the simple models considered here may only help to develop
tools and analysis to understand more realistic systems for which it
is then necessary to consider the role of merging, of dissipation and
the fact that they cannot be simply considered isolated. Whether these
complications qualitatively change the evolution with respect to the
simple systems considered here will be studied in a forthcoming
publication.  However, the collapse of an isolated cloud of
self-gravitating particles may describe the formation of elliptical
galaxies \citep{henon_1964,vanalbada_1982,paper1} which are
characterised by the de Vaucouleurs $R^{1/4}$ law. A possible
analytical approximation to the de-projected $R^{1/4}$ is given by the
Hernquist profile \citep{hernquist1990,visbal_2012} which is indeed
characterised by a $r^{-4}$ decay at large radii.

The paper is organised as follows.  We present in
Sect.\ref{simulations} the details of the numerical simulations that
we have performed. The case of an initially scale free profile is
discussed in Sect.\ref{scalefree}. We then consider in
Sect.\ref{other} more complex density profiles, which have a
characteristic length. Finally in Sect.\ref{discussion} we discuss the
results draw our main conclusions.


\section{The simulations} 
\label{simulations}

The initial conditions are generated by distributing $N$ particles
randomly with density $n(r)$, where $r$ is the distance from the
origin of coordinates. For the scale free density profiles
$n(r)=n_\alpha r^{-\alpha}$, where $0 \le \alpha < 3$: for larger
values of $\alpha$ and $r \rightarrow 0$ there are too many nearest
neighbours with arbitrarily small distance. In such a situation an
accurate numerical integration of the equations of motion becomes very
problematic. We put a lower and an upper cut-off: the latter is of the
order of the softening length of the code and the former, $R_0$ , is
larger than any characteristic scale of the profile (if present).

We generate a series of spherical clouds of particles, with $N=10^4$
and with different initial virial ratio $b_0=b(t=0)=2K_0/W_0$ (where
$K/W$ is the kinetic/potential energy).  In order to assign velocities
we take the velocity components to have a uniform probability density
function (PDF) in the range $[-V_0,V_0]$ so that the modulus of the
velocity is constrained to be in a sphere of radius $V_0$.  The
velocity PDF is thus
\be 
g(v) = \frac{3}{V_0^3} v^2 \;\; \mbox{for}
\;\; v \le V_0 
\ee 
and zero otherwise. 
The mean square value of the velocity is 
\be 
\label{v02}
\langle v^2 \rangle = \int_0^{V_0} v^2 g(v) dv = \frac{3}{5}V_0^2
=  \frac{W_0 b_0}{mN} \;.
\ee

We used the publicly available tree-code GADGET
\citep{gadget_paper,springel_2005} to run the simulations.  We have
chosen the same combination of numerical parameters as in
\citet{paper1}, to which we refer the reader for further details on
the numerical issues. 


\section{Scale-free profiles}
\label{scalefree}

We first discuss the relaxation of a spherical and isolated cloud,
with a with a power-law profile, initially at rest. Then we 
consider the effect of a non-zero initial velocity dispersion.

\subsection{Virial ratio} 

\begin{figure}
\vspace{1cm}
{
\par\centering \resizebox*{9cm}{8cm}{\includegraphics*{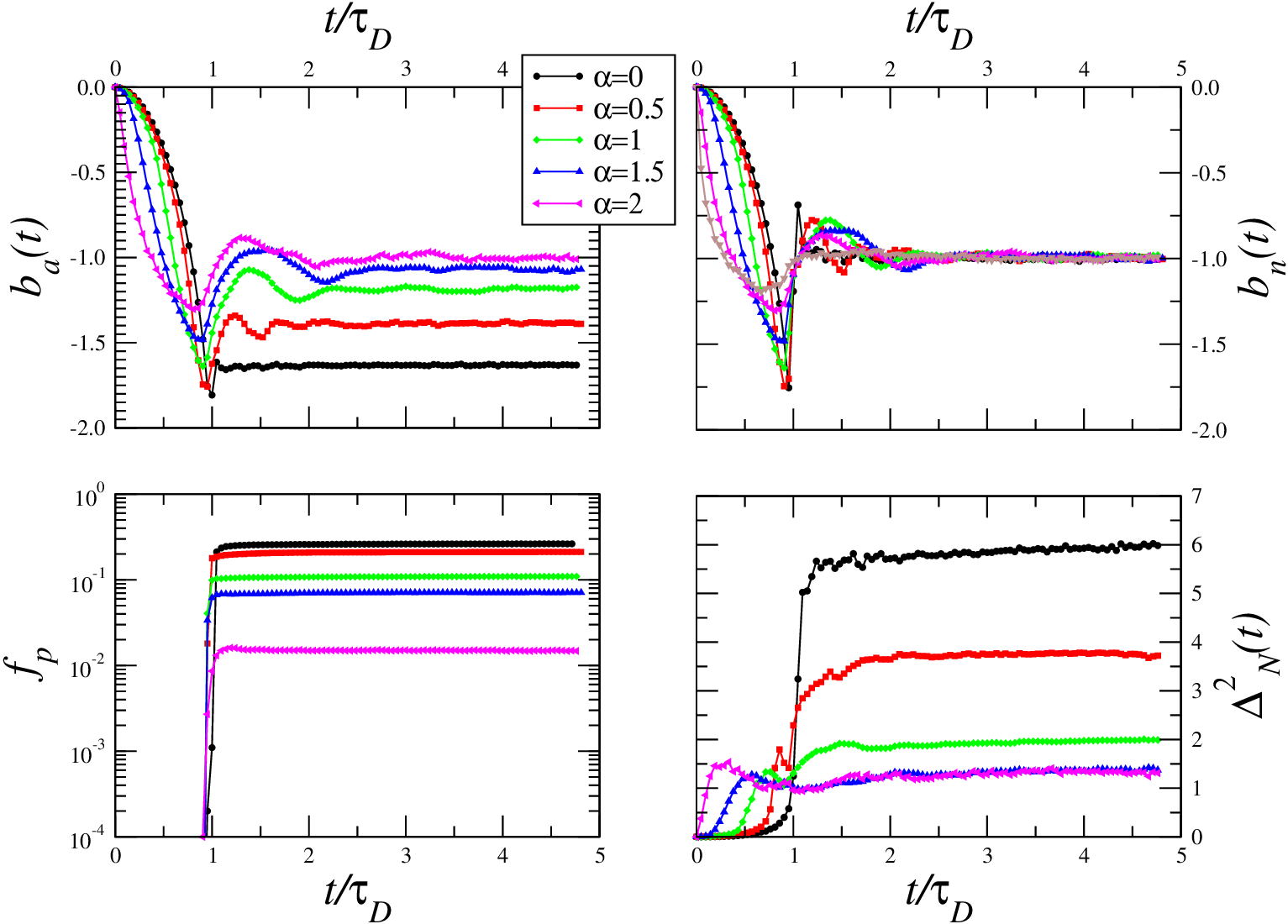}} 
\par\centering
}
\caption{ Upper left panel: Virial ratio of all particles.  Upper
  right panel: Virial ratio of particles with negative total energy.
  Bottom left panel: Fraction of particles with positive energy.
  Bottom right panel: Behaviors of the (normalised) energy exchanged
  $\Delta_N(t)$ (see text for details).}
\label{fig:fig1a} 
\end{figure}

  The virial ratio for all system particles $b_{a}(t)$ and the virial
  ratio for bound $b_{n}(t)$ have a different amplitude (see upper
  left and right panels of Fig.\ref{fig:fig1a}): this can be easily
  explained as due to the fact that part of the particles have been
  ejected during the collapse. Indeed, the fraction of particles
  $f_p(t) $ with positive energy has an abrupt change becoming larger
  than zero for $t \ge \tau_D$ (see Fig.\ref{fig:fig1a} --- bottom
  left panel). For $\alpha \ge 2$ only $f_p(t) \approx 1 \%$ of the
  particles gain positive energy during the collapse while for
  $\alpha=0$ we find $f_p(t) > 20\%$: the amount of ejected particles
  decreases with $\alpha$ and so the difference between $b_{tot}(t)$
  and $b_{neg}(t)$.



\subsection{Ejection}
\label{ejection} 

As discussed in detail in \citet{paper2,paper1} the mechanism allowing
particles to gain energy during the collapse phase arises from the
interplay between the growth of density perturbations in the
collapsing cloud and the finite size of the system. In practice,
particles that were originally in the outer shells of the cloud
arrive later than the others in the centre, thus moving, for a short
time interval around $\tau_D$, in a rapidly varying gravitational
potential.  As noticed above, even when $\alpha>0$ some particles
acquire positive energy at about $t \approx \tau_D$, i.e. when all
other particles have already arrived to the centre and inverted their
velocities. The difference between the case $\alpha=0$ and $\alpha>0$,
and so the variation of $f_p$ with $\alpha$, is due to the different
time of arrival at the centre of particles initially placed at
different radii.

In the fluid limit and neglecting density perturbations, 
a particle initially placed at $r_0$ obeys to the
following parametric equations:
\bea
&& 
r(\xi) = \frac{r_0}{2} (1 + \cos(\xi))
\\ \nonumber 
&&
t(\xi) = \sqrt{\frac{r_0^3}{8GM(r_0)}} \left( \xi + \sin(\xi) \right)
\;, 
\label{scm2}
\eea 
where 
\be
M(r_0)= \int_0^{r_0} m n_\alpha r^{-\alpha} 4 \pi r^2 dr = 
\frac{4 \pi }{3- \alpha} m n_\alpha r_0^{3-\alpha} \; . 
\ee
Thus we get that a particle initially at $r_0$ arrives at the origin
for $\xi=\pi$ and thus at
\be
\label{taur0}
\tau(r_0;\alpha) = 
\sqrt{\frac{ r_0^3 \pi^2}{8 G M(r_0)}}
=
\sqrt{\frac{ (3-\alpha) \pi r_0^\alpha}{32 m n_\alpha G}}\;, 
\ee
i.e., particles initially placed in the outer shells arrive later
than the others for $\alpha>0$ even if they follow their unperturbed
trajectories. Instead, in this same limit, for $\alpha=0$ all
particles should arrive simultaneously at the
centre, i.e., $\tau(r_0,0) \equiv \tau_D \;\; \forall r_0$.

\begin{figure}
\vspace{1cm}
{
\par\centering \resizebox*{9cm}{8cm}{\includegraphics*{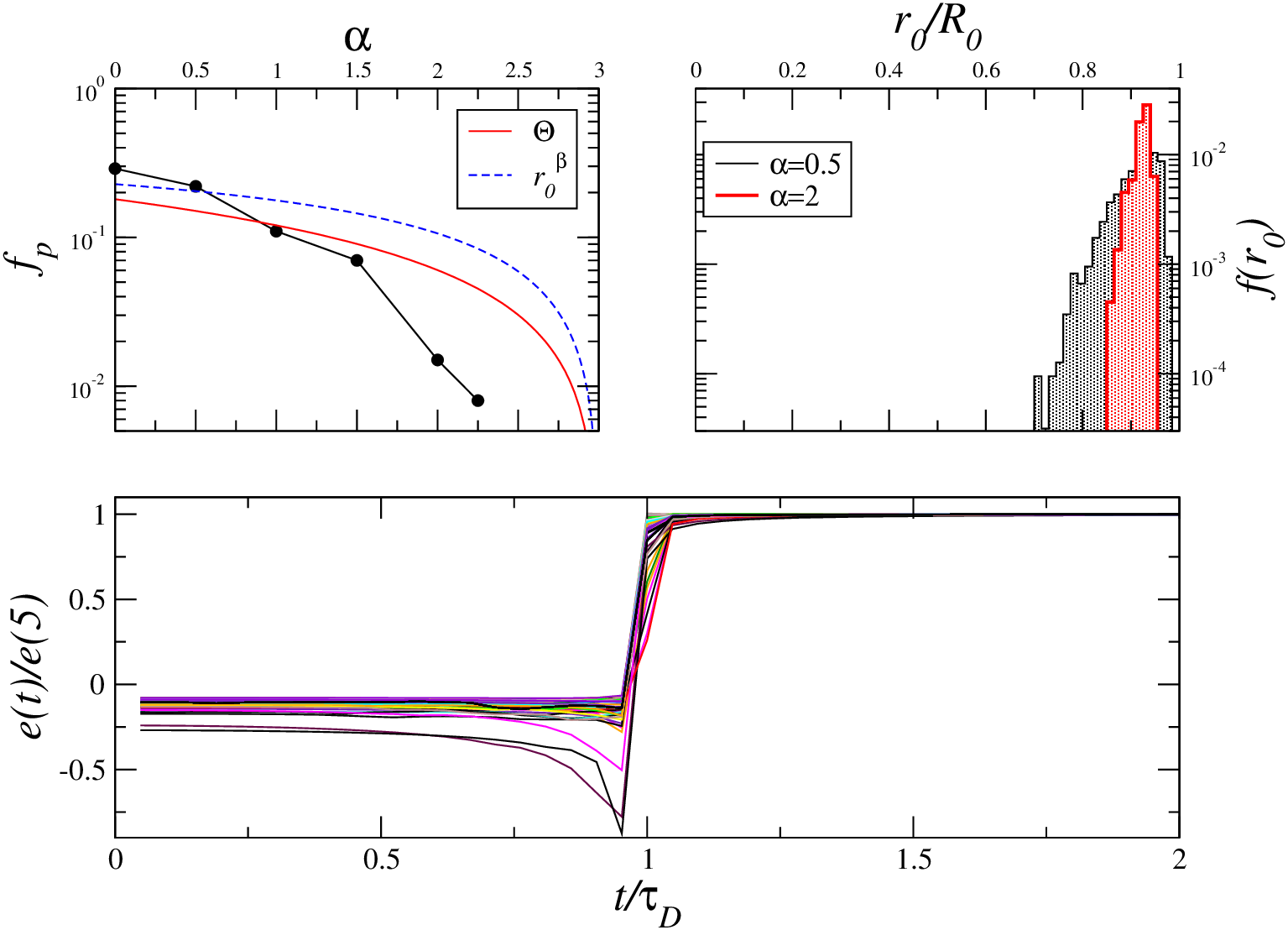}} 
\par\centering
}
\caption{Upper left panel: Number of ejected particles as a function
  of their initial radial position for $\alpha=0.5$ and $\alpha=2$. B
  Upper right panel: Fraction of the particles with positive energy as
  a function of $\alpha$. The dashed line corresponds to a radial
  probability of ejection described by the Heaviside step function
  (i.e., $p(r_0)=\Theta(r_0 -R_0 + \Delta R)$ --- see
  Eq.\ref{fp_alpha}) while the solid line to a power-law radial
  probability $p(r_0) \sim r_0^\beta$ with $\beta =4$
  (i.e. Eq.\ref{fp_alpha2}).  Bottom panel: Behaviour of the energy
  $e_p(t)$ for some of the ejected particles as a function of time,
  normalised to the energy at $t=5 \tau_D$ and for $\alpha=0$.}
\label{fig:fig1b} 
\end{figure}

Noticing that for $r_0=R_0$ we have $M(R_0)=mN$ and
$\rho_0=M(R_0)/V(R_0)= 3 mN/(4\pi R_0^3)$ we find that
\be
\label{taur0b}
\tau(r_0<R_0;\alpha) < \tau(R_0;\alpha) = \tau_D  \;\; \forall \alpha \;.  
\ee 
Thus the maximum time taken by a particle to arrive to the centre
does not depend on $\alpha$.  Given that we detect particles with
positive energy only for $t \ge \tau_D$, being $\tau_D$ the time
taken by the particles in the outer shell to arrive at the centre, we
may conjecture that, as it occurs for the $\alpha=0$ case, ejected
particles are those which initially lie in the outer shells: indeed
Fig.\ref{fig:fig1b} (upper right panel) shows that this is the case.

In order to estimate the scaling of the fraction of ejected particles
as a function of $\alpha$, let us suppose that all ejected particles
belong initially to the outer shell, i.e. they have at time $t=0$
radius  $r_0 \in [R_0 - \Delta R, R_0]$.  The number of
particles in this last shell scale as
\footnote{This corresponds to assuming a radial probability
  distribution for a particle initially placed at $r_0$ for being
  ejected at $t>\tau_D$ of the type $p(r_0)= \Theta(r_0 -R_0 + \Delta
  R)$ where $\Theta(x)$ is the Heaviside step function.}
\be
\label{fp_alpha} 
\Delta N = \int_{R_0 -\Delta R}^{R_0} 4\pi
n_\alpha r^{2-\alpha} dr = N \left( 1 - \left( 1 - \frac{\Delta
  R}{R_0} \right)^{3-\alpha} \right) 
\ee
where we used the normalisation condition ($N$ is fixed) 
\be
n_\alpha = \frac{(3-\alpha)N}{4 \pi R_0^{3-\alpha}}\;.
\ee
The scaling behaviour of the fraction of ejected particles as a
function of $\alpha$ derived from Eq.\ref{fp_alpha}, i.e. $f_p= \Delta
N/N$, is shown in Fig.\ref{fig:fig1b} (upper left panel), together
with the simulation points.  (Note that the amplitude is fixed by the
choice of $\Delta R$.)  One may see that a very rough agreement is
found: in principle, one should use the true probability
$p(r_0)$ for a particle to be ejected as a function of its initial
radial position $r_0$.  For instance, by assuming $p(r_0) \sim
r_0^\beta$ with $\beta >3$, as measured in \cite{paper2} for the case
$\alpha=0$, we find
\be
\label{fp_alpha2} 
\frac{\Delta N}{N} \propto \frac{3-\alpha}{3-\alpha+\beta} \;. 
\ee
Even in this case the agreement is very rough: however the
assumption that $p(r_0)$ has the same exponent for all values of
$\alpha$ is probably too strong. Indeed, as mentioned above, the
physical mechanism of ejection is due to a complex interplay between
the fluctuations present in the distribution and the finite size of
the system. 

\subsection{Energy exchange} 
\label{enex}

It is interesting to consider how individual particles gain energy as
they pass through the core. In Fig.\ref{fig:fig1b} (bottom panel) it
is plotted the behaviour of the energy $e_p(t)$ for some of the
particles that will be ejected after the collapse as a function of
time, normalised to the energy at late times (i.e. $t=5 \tau_D$) and
in the case $\alpha=0$. One may note that there is a single abrupt
change of the energy at $t\approx \tau_D$, a fact that shows that
particles are being accelerated by large scale changes of the mean
field potential, instead of getting rapid kicks from encounters with
individual particles.

A statistical manner to monitor how particles energy changes during
time consists in measuring the quantity \citep{paper1}
\be
\label{delta2}
\langle \Delta^2 (t) \rangle = \frac{1}{N(N-1)} 
\frac{\sum_{i,j=1 \; i\ne j }^{N} (e_p^i(t) -e_p^j(t))^2 }
{\langle e(t) \rangle^2}
\ee
where $e_p^i(t)$ the average energy per particle is defined as 
\be
\langle e(t) \rangle = \frac{\sum_{i=1}^{N_t} e_p^i(t)}{N} \;. 
\ee 
Fig.\ref{fig:fig1a} (bottom right panel) shows the behaviour of $
\Delta^2_N(t) = \langle \Delta^2 (t) \rangle-\langle \Delta^2 (0)
\rangle$ in the various simulations: this changes with time
when particles exchange energy, i.e. when the particle normalised
energy distribution changes.

It is interesting to note that while for $\alpha=0$ the exchange of
energy occurs simultaneously with the strongest phase of the collapse
at $t\approx \tau_D$, for $\alpha>0$ particles exchange energy also
for $t<\tau_D$. This occurs because, for $\alpha>0$ the $i^{th}$
particle, initially located at $r_0^i > r_0^j \gg 0$, arrives at the
centre, if it follows its unperturbed trajectory, later than the
$j^{th}$ one (see Eq.\ref{taur0}). Thus when the $i^{th}$ particle
arrives at the centre, inverting its velocity, it moves in a varying
potential field generated by all other particles which were initially
placed at distance $< r_0^i$.  However the exchange of energy, as long
as $r_0 < R_0$, is not enough to allow a fraction of the particles to
escape from the system.  This is because the gain of kinetic energy
depends on the deepness of the potential well and thus on the total
mass generating it \citep{paper2}.

\begin{figure}
\vspace{1cm}
{
\par\centering \resizebox*{9cm}{8cm}{\includegraphics*{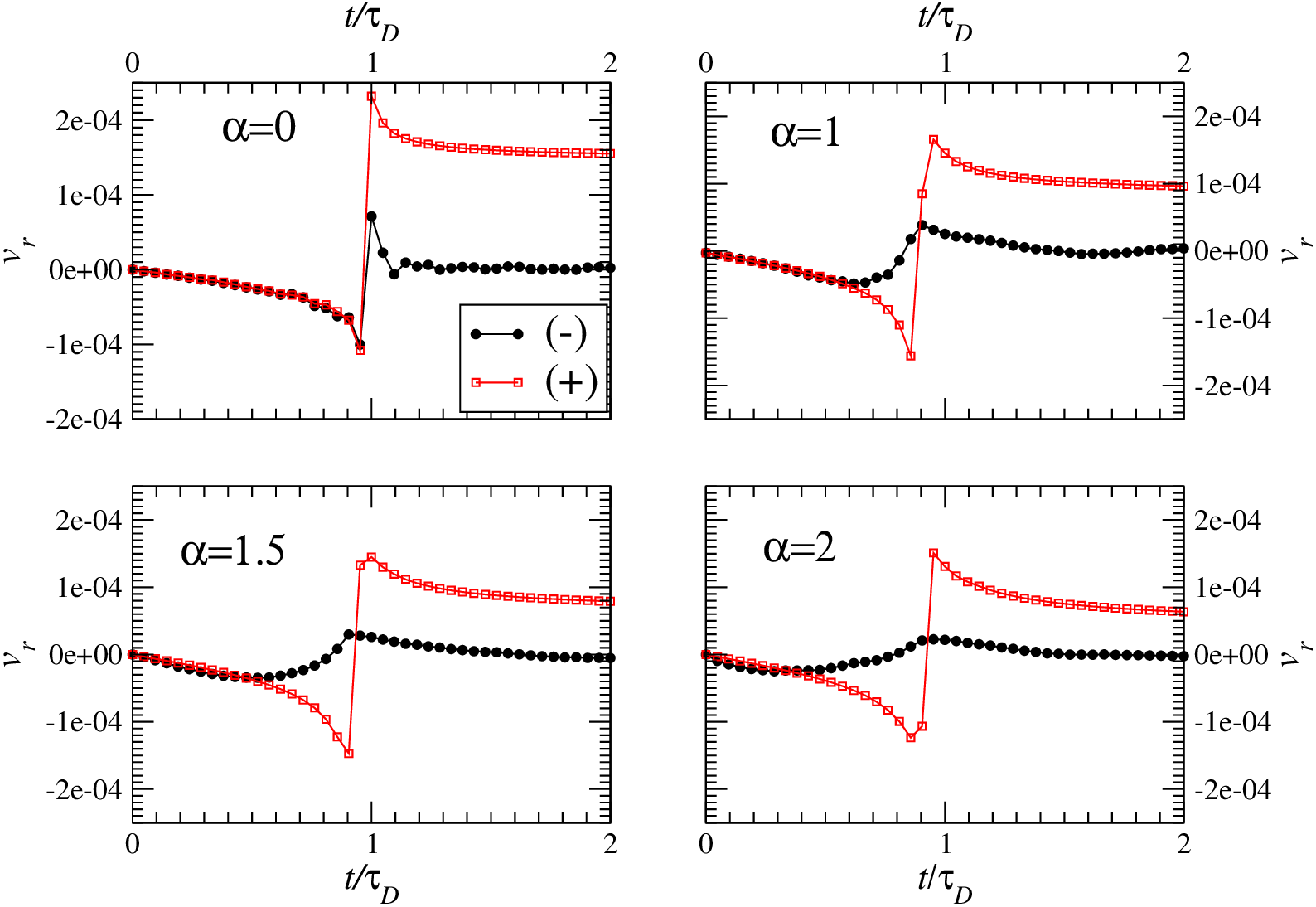}} 
\par\centering
}
\caption{Behaviour of the average radial velocity for particles that
  at $t = 5 \tau_D$ are bound (-) and ejected (+).}
\label{radial_velocity}
\end{figure}

Finally, we have plotted in Fig.\ref{radial_velocity} the average
radial velocity at time $t$ of the particles that at later times,
i.e. at $5 \tau_D$, have positive (+) and negative energy (-).  One
may notice that the change in the radial velocity direction occurs
sooner for the latter particles than for the former ones. Particles
that will be ejected arrive later than all others, having about the
same velocity independently on $\alpha$. Indeed, as they belong to the
outer shell, they are attracted by the same gravitational field. For
this reason the dynamics of the ejection is the same for different
values of $\alpha$: what changes is, as previously discussed, only the
amount of mass and energy that is ejected from the system.



\subsection{Density profile}

\begin{figure}
\vspace{1cm}
{
\par\centering \resizebox*{9cm}{8cm}{\includegraphics*{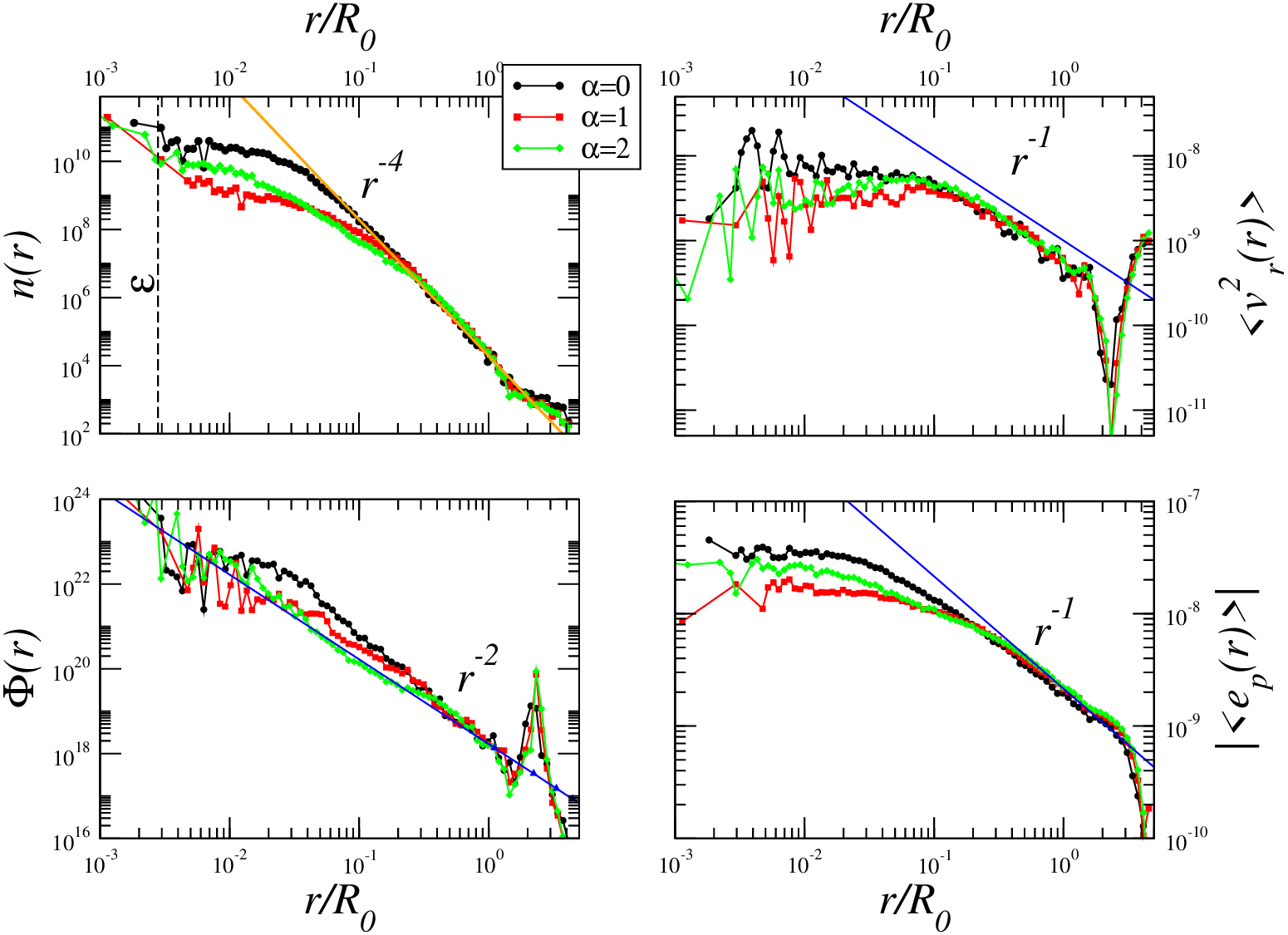}} 
\par\centering
}
\caption{Upper left panel: Density profiles $n(r)$ at $t=5 \tau_D$:
  the amplitude for $\alpha>0$ has been arbitrarily normalised to have
  the same large radii amplitude of $n(r)$ for $\alpha=0$.  Upper
  right panel: Square value of the radial component of the velocity
  $\langle v^2_r(r) \rangle$ as a function of the radius at $t=5 \tau_D$
  for different values of $\alpha$.  Bottom left panel: Pseudo
  phase-space density $\Phi(r)$ as a function of the radius at $t=5
  \tau_D$ for different values of $\alpha$: the different curves have
  been normalised to have same amplitude at large radii as the
  $\alpha=0$ case.  Bottom right panel: Average energy per particle
  $\langle e_p(r) \rangle$ (absolute value) as a function of the
  radius and for different values of $\alpha$.  }
\label{fig:fig2}
\end{figure}

In the case of an initially uniform density profile, i.e. $\alpha=0$,
the radial density profile of the virialized structure formed by bound
particles after the collapse was found to have a $r^{-4}$ decay.
\citep{paper2,paper1}.  For $\alpha >0$ one may notice (see
Fig.\ref{fig:fig2} --- upper left panel) that at small radii the
different profiles have different slopes which are close to the
initial ones (for a value of the radius of order of the force
softening length $\varepsilon$ the profile seems to flatten, probably
due to the fact that the interaction is smoothed) A simple
interpretation of this behaviour is the following: particles in the
dense core were already very strongly bound at $t=0$, as the density
has a cusp for $\alpha>0$ by construction; thus the energy exchange
during the violent phase of the collapse represents a small
perturbation that is not able to alter significantly the inner
particles trajectories.

On the other hand for $r/R_0 \approx 1$ all profiles converge to the
same shape, i.e.  $n(r) \sim r^{-4} \;\; \forall \alpha$. We thus
conclude that the dynamical mechanism underlying the formation of the
$r^{-4}$ should be the same for all $\alpha$ values, i.e.  it is
related to process giving rise to mass and energy ejection.

A simple way to focus on the properties of the $r^{-4}$ tail is to
monitor how this changes during time evolution. Indeed, as mentioned
above, in the case $\alpha=0$, the large scale tail which decays as
$n(r)\sim r^{-4}$ is made of particles with almost zero energy
orbiting around the dense core at smaller radii. As time passes these
particles may reach larger and larger radii and thus the tail extends
to larger and larger radii. This is indeed what is observed not only
in the case $\alpha=0$ but also for $\alpha>0$. This evidence
corroborate the conclusion that the density profile tail properties
have the same physical origin for all $\alpha$ values.  We will come
back on this point in what follows.


\subsection{Radial velocity dispersion profile}

In Fig.\ref{fig:fig2} (upper right panel) it is shown the mean square
value of the radial velocity as a function of the radius 
$\sigma^2_r(r) = \langle v^2_r(r) \rangle$ at $t=5\tau_D$ for
different values of $\alpha$. One may note that the different curves
are almost indistinguishable and that at large radii, i.e. $r/R_0
>0.1$, they show an approximated Keplerian $\sigma^2_r(r)\sim r ^{-1}$
behaviour.

The time evolution of $\langle v^2_r(r) \rangle$ is again very similar
for different values of $\alpha$ and compatible with the physical
model presented in \citet{paper1}: a few particles with almost zero
energy continue to increase their distance from the structure core,
remaining however bound. Note that, at a given time, the break of
the $r^{-1}$ behaviour is found at the same radii for different values
of the $\alpha$: this reflects the fact that particles belonging to
the tail not only obey to the same equation of motions but also are
subjected to the gravitational effect of approximately the same core
mass.

\subsection{Pseudo phase-space density} 

The behaviour pseudo phase-space density $\Phi(r)=n(r)/\sigma_r^3(r)$
(see Fig.\ref{fig:fig2} --- bottom left panel) is a combination of the
density and radial velocity dispersion profiles.  In agreement with
the previous results, at large enough radii it shows the same
behaviour independently on $\alpha$; indeed, a fit with a function of
the type $\Phi(r) \sim \frac{r^{3/2}}{r^4} \sim r^{-5/2}$ describes
only the large scale tail. Finally we note that, as for the radial
velocity dispersion, at a given time, the break of the $r^{-5/2}$ tail
is found at the same scale for different values of $\alpha$ (in this,
at $t=5\tau_D$, there is a bump at $r/R_0\approx 3$).


\subsection{Energy distribution} 

\begin{figure}
\vspace{1cm} { \par\centering
  \resizebox*{9cm}{5cm}{\includegraphics*{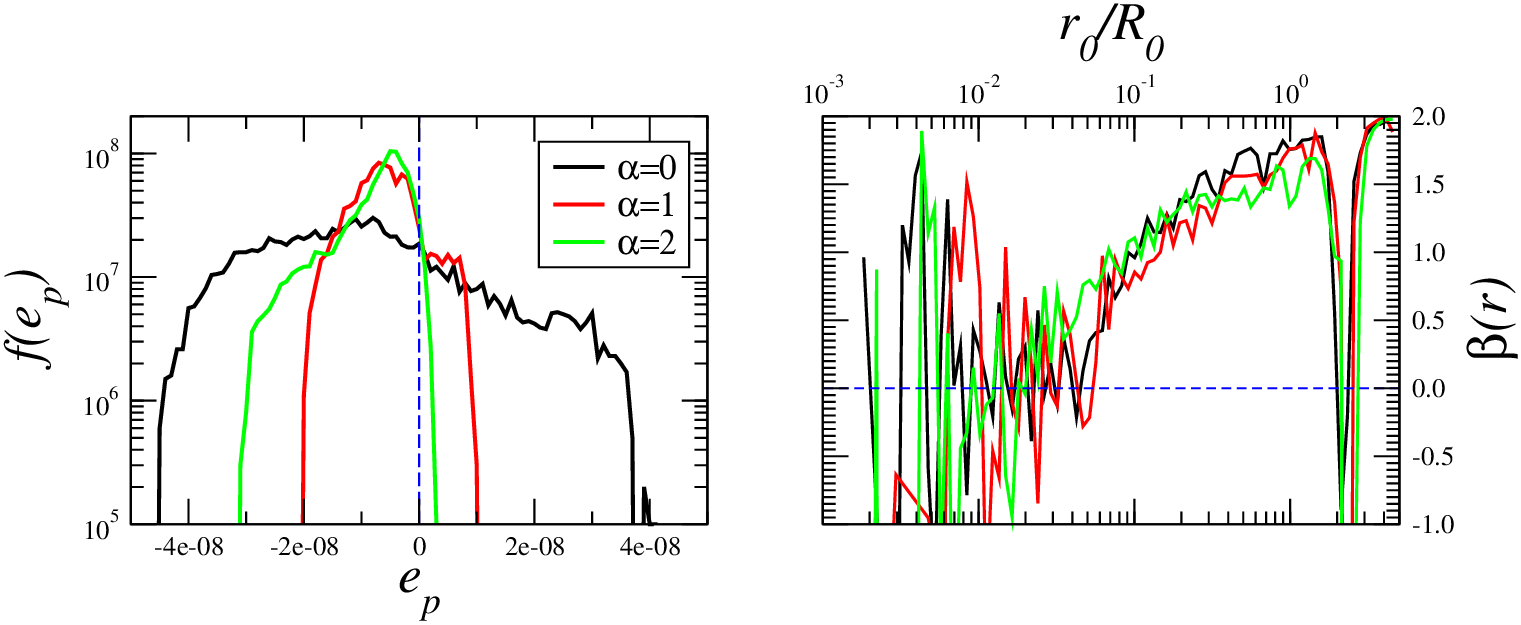}}
  \par\centering }
\caption{Left panel: Particle total energy distribution for different
  values of $\alpha$ Right Panel: Velocity anisotropy at $t=
  5\tau_D$. When $\beta(r)>0$ the velocity dispersion is dominated by
  its radial component. }
\label{fig:fig2b}
\end{figure}

The distribution of particles total energy at $t\approx 5\tau_D$,
i.e. when the virialized structure is already formed, is shown in
Fig.\ref{fig:fig2b} (left panel).  One may note that the larger is $\alpha$
the smaller is the fraction of particles with positive energy, a
result clearly in agreement with the ejection mechanism discussed in
Sect.\ref{ejection}.

The average energy of a particle with radial distance $r$ from the structure
centre is displayed in Fig.\ref{fig:fig2} (bottom right panel). At
large radii, i.e. $r> 1/10R_0$, the energy displays a $r^{-1}$
behaviour, in agreement with the fact that the gravitational potential
energy is dominated by the inner core and thus that the velocity
dispersion presents a Keplerian decay.  On the other hand, at small
radii, the different density profile give rise to slightly different
potential and kinetic energy profiles.


\subsection{Velocity anisotropy}

Fig.\ref{fig:fig2b} shows the behaviour, in the different
simulations, of the velocity anisotropy parameter defined as 
\be
\label{alpha} 
\beta(r) = 2 - \frac{\overline{v_t(r)}^2} {\overline{v_r(r)}^2 } \;,
\ee 
where $v_t(r)/v_r(t)$ is the velocity in the transversal/radial
direction.  In the outer regions of the virialized structure where the
density scales as $n(r) \sim r^{-4}$, radial orbits dominate the
motion, as it was already noticed by
\citet{vanalbada_1982,trenti2005,paper1} for the $\alpha=0$ case.  It
is interesting to note that not only the behaviour of $\beta(r)$ is
essentially the same for different values of $\alpha$, but also its
amplitude. This, together with the evidences already discussed from
the radial velocity dispersion and its time dependence, confirms the
fact that the particle dynamics in the large scale tail is the same for all
values of $\alpha$ considered.


\subsection{The case of a nonzero initial velocity dispersion} 
\label{Section2} 

For an initially uniform cloud (i.e., $\alpha=0$) and for warm enough
initial velocity dispersion ejection of mass and energy does not occur
anymore. In \citet{paper1} we found that the critical value of the
initial virial ratio was $b_0^c \approx -1/2$: for $b_0 > b_0^c$
ejection occurs and for $b_0 < b_0^c$ it does not occur anymore. The
value of $b_0^c$ is approximate as it was conjectured to depend on the
number of points $N$ used in the simulation. The reason for not having
ejection anymore is due to the fact that an high enough velocity
dispersion cause shell crossing well before the collapse time
$\tau_D$. Such a shell crossing erase the temporal lag of the outer
shell, thus avoiding that particles initially placed at the outer
boundaries of the system arrive later, then moving in a rapidly time
varying gravitational field.

A similar mechanism should be working also for $\alpha>0$.  For each
value of $\alpha$ and for each $N$ one should be able to find a
critical value of the initial virial ratio,
i.e. $b_0^c=b_0^c(N;\alpha)$. Here we have limited our study to
$b_0=-1/2,-1$ and $\alpha=1,2$ for which we have not found mass and
energy ejection. Instead of determining the precise value of $b_0^c$
we are interested here in the shape of the density and velocity
profiles when ejection does not occur.

\begin{figure}
\vspace{1cm} { \par\centering
  \resizebox*{9cm}{8cm}{\includegraphics*{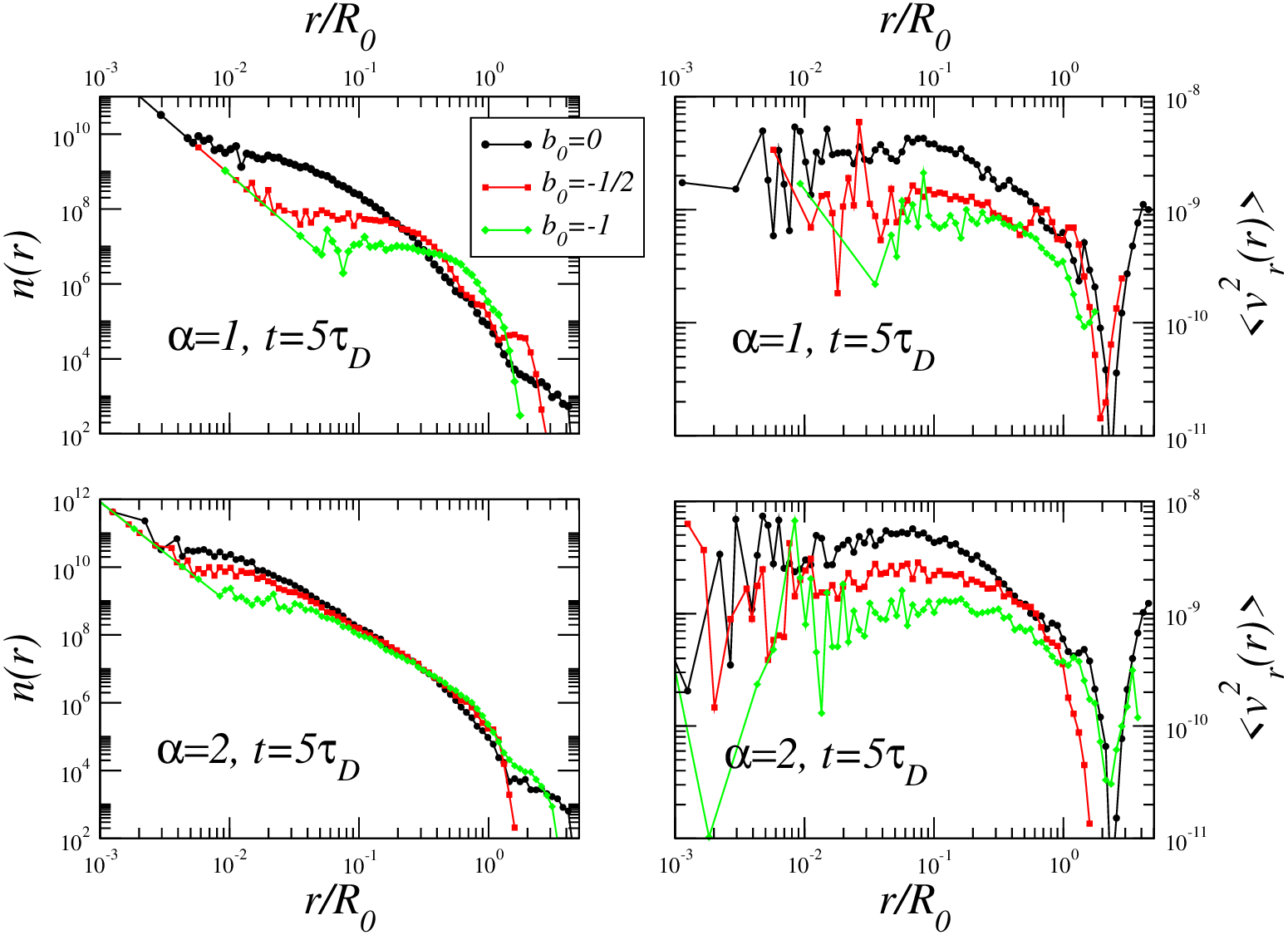}}
  \par\centering }
\caption{Density profile and  velocity dispersion 
for $\alpha=1$ (upper panels)  and $\alpha=2$ 
(bottom panels) and $b_0=0, -1/2, -1$ at $t=5\tau_D$.}
\label{DenProf_alpha1}
\end{figure}

In this case the collapse is more gentle, and the virial ratio shows a
series of damped oscillations which drive the system toward the
quasi-equilibrium state.  In Fig.\ref{DenProf_alpha1} we show
respectively the density profile, velocity dispersion and pseudo
phase-space density for $\alpha=1,2$ and $b_0=0,-1/2,-1$.  One may
note that (i) the smaller is $b_0$ and the smaller the difference
between the input and output density profile.  (ii) For different
values of $b_0$ there is a difference in the behaviour of all density
and velocity profiles, a fact that points toward an important role of
the initial conditions.  (iii) The case $\alpha=2$ presents the
smaller change in the density profile $n(r)$ between the initial and
final condition: the larger is $\alpha$ and the deeper the potential
well at small scales and thus the smaller the perturbation introduced
by a given amount of kinetic energy.
We thus conclude that when the collapse is violent, for the effect on an
initial velocity dispersion, there is a clear dependence of the
virialized structure properties on the initial conditions.


\section{Relaxation of more complex density profiles} 
\label{other}
In this section we have considered and initially perfectly cold cloud
of self-gravitating particles, whose profile is described by:
(i) the Plummer sphere
\be
\label{plummer}
n(r) = \frac{A}{\left(1+\frac{r^2}{r_s^2}\right)^{\frac{5}{2}}} \;,
\ee 
(ii) the Hernquist profile ($\gamma=3$), (iii) the NFW profile
($\gamma=2$) \be
\label{hernquist} 
n(r) = \frac{A}{r \left( 1+\frac{r}{r_s}\right)^{\gamma}} \;,
\ee
and the (iv) Gaussian profile
\be
\label{gaussian} 
n(r) = A \exp\left(-\left(\frac{r}{r_s}\right)^{2}\right) \;, 
\ee
where in Eqs.\ref{plummer}-\ref{gaussian} $A$ and $r_s$ are two
constants that are fixed by the number of points $N_p$ initially
belonging to the cloud and by the requirement that the radius enclosing
half of the mass is the same for all profiles, so that the collapse
time is about the same in the different cases.

\begin{figure}
\vspace{1cm}
{
\par\centering \resizebox*{9cm}{8cm}{\includegraphics*{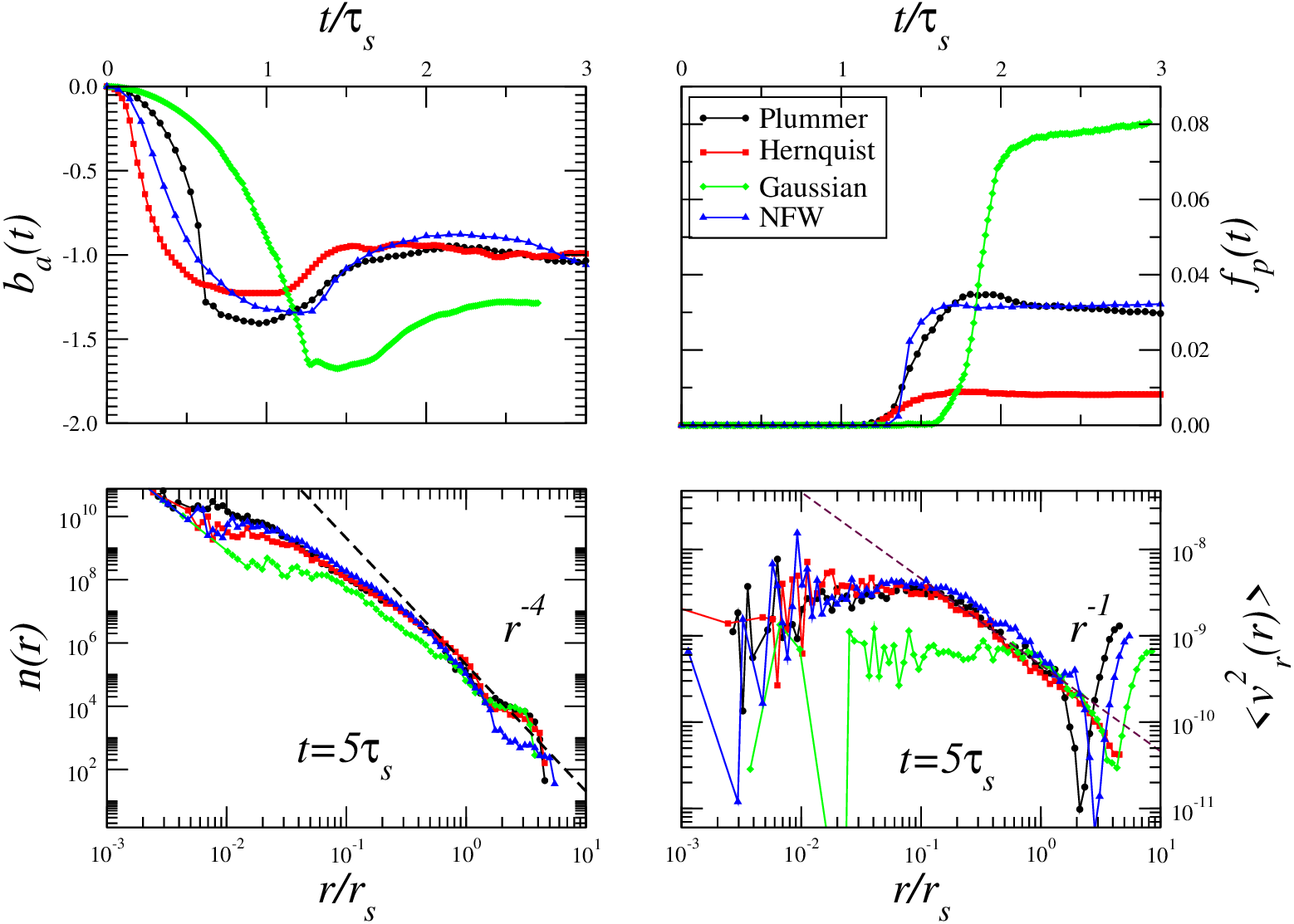}} 
\par\centering
}
\caption{Upper left panel: Virial ratio of all particles.  Upper right
  panel: Fraction of particles with positive energy.  Bottom left
  panel: density profiles $n(r)$ at $t=5 \tau_D$ Bottom right panel:
  radial velocity dispersion $\langle v^2_r(r) \rangle$ at $t=5
  \tau_D$ }
\label{fig:fig5c} 
\end{figure}

Results are shown in Fig.\ref{fig:fig5c}: we have normalised the time
to the characteristic time $\tau_s$ for collapse for each density
profile and the radius to $r_s$ defined in
Eqs.\ref{plummer}-\ref{gaussian}. One may note that there are
similarities with the behaviours found in
Figs.\ref{fig:fig1a}-\ref{fig:fig2}. In particular in all case it is
observed ejection of a fraction of the particles and the formation of
a $n(r) \sim r^{-4}$ density profile and a radial velocity dispersion
decaying as $r^{-1}$. Qualitatively the dynamics is thus similar to
the case discussed in the previous section.

\section{Discussion and conclusions} 
\label{discussion} 

The collapse of an isolated spherical cloud of self-gravitating
particles represents a paradigmatic example of the dynamics giving
rise, through a global relaxation process, to a virialized structure
in a quasi-equilibrium state, i.e.  a state that has a life-time that
diverges with the number of system particles and that is far from
thermodynamical equilibrium.  If the collapse is violent enough part
of the initial mass and energy of the cloud is ejected in the form of
free particles. This ejection occurs, for the case of an initially
uniform cloud, when the initial velocity dispersion is small enough
\citep{paper1}. In this case some non-trivial density and velocity
profiles have been observed: (i) the radial density decays as
$n(r)\sim r^{-4}$ at large enough radii, (ii) the radial velocity
dispersion decays as $\sigma^2_r \sim r^{-1}$ and thus (iii) the
pseudo phase-space density decays as $n(r)/\sigma_r^3(r) \sim
r^{-5/2}$ \citep{paper1}.  All these behaviors can be understood in
the framework of a simple dynamical model in which the bound
particles that have the largest energies, belong to the tail of the
profile and are orbiting in a central gravitational potential
generated by the inner dense core \citep{paper1}.

The problem that we have considered in this paper concerns whether the
same density and velocity profiles are observed when the initial
conditions are not represented by a uniform cloud of particles. In
particular, we have firstly considered an initial density profile of
the type $n(r) \sim r^{-\alpha}$ with $0 \le \alpha <3$.  When the
initial velocity dispersion is zero, the situation for $\alpha>0$ is
similar to the uniform case, i.e. $\alpha=0$, in various
aspects. During the collapse phase a fraction of the particles gain
enough kinetic energy to escape from the system, i.e.  there is
ejection of mass and energy.  The virialized structure formed after
the collapse presents, at large enough radii, the same behaviors
(i)-(iii) discussed above for the initially uniform cloud.  In
addition, radial orbits dominate the motion of particles in the
density profile tail in all cases.  The time evolution of these
quantities shows that the underlying dynamics determining the large
radii properties of the virialized structures is identical to the
$\alpha=0$ case.  These common features are thus generated by the
dynamical mechanism acting during the collapse, and responsible of the
mass and energy ejection. For this reason they are independent on the
initial properties of the self-gravitating cloud.

On the other hand, when the initial velocity dispersion is high
enough, we find that there is no ejection of mass and energy so that
the density and velocity profiles do not present any power-law
behaviour at large radii. In this case the collapse is more gentle and
consists in a few oscillations that drive the system to a
quasi-equilibrium configuration and the virialized structure displays
features that depend on the initial conditions.

In addition, we have considered other initial density profiles which
are not scale-free, as the Plummer, the Hernquist, the NFW and the
Gaussian profile. In all these cases we found that, when the initial
velocity dispersion is set to zero, the density profiles converges,
for large enough radii, at the same behaviors (i)-(iii) discussed
above. These results are in agreement with \citet{visbal_2012} who
have recently found that a Plummer sphere, originally at rest,
collapses, in simulations where non-radial motions are both present or
suppressed, forming a virialized state with a $r^{-4}$ tail at large
radii. Note that a simple explanation for the formation of the
$r^{-4}$ was proposed by \citet{white_1987,jaffe_1987}; in
\citet{paper1} we introduced a more detailed model of particle
dynamics, which considers as well the behaviour of the bound particles
with almost zero total energy, and a series of analyses that
corroborate the hypothesis that particle with very small binding
energies form the $r^{-4}$ tail.

We thus conclude that when a cloud undergoes a violent collapse ,
i.e. when part of its initial particles gain energy during the
collapse to eventually escape as free ones, it reaches a virialized
quasi-equilibrium configuration that displays properties which are
formed by the dynamical relaxation mechanism. In particular, these
features characterise the tail of the density and radial velocity
dispersion profiles and they are related to the behaviour of those
bound particles which were initially in the outer shells of the
distributions. 

Instead, strongly bound particles, i.e. those in the central dense
core, have properties which depend on the initial conditions. These
are the particles which already initially were strongly bound, given
the cusp of the density profile: for this reason the energy exchange
during the violent phase of the collapse represents a small
perturbation that is not able to alter significantly their
trajectories.

As a final remark we note that the systems considered here, although
pose a series of interesting theoretical questions, differ for many
important aspects not only from real systems but also from structures
formed in cosmological N-body simulations. In particular, a more
detailed study of the modification of the isolated system dynamics due
to space expansion, continuous mass accretion, merging with macro
objects and tidal interactions with background structures and a
careful analysis of numerical and finite size issues \cite{jsl12} is
necessary to clarify the relation between the simple self-gravitating
clouds considered in this paper and halos formed in cosmological
simulations.

I thank Michael Joyce for useful discussions and comments, Roberto
Ammendola and Nazario Tantalo for the valuable assistance in the use
of the Fermi supercomputer where the simulations have been performed.



\bibliographystyle{mn2e}

\end{document}